\documentclass[12pt,twoside]{article}
\usepackage[mathscr]{eucal}
\usepackage{amsmath,amsfonts,amssymb,amsthm,mathabx,empheq}
\usepackage{times}
\usepackage{pdfsync}
\usepackage{cite}
\usepackage{url}
\usepackage{hyperref}
\usepackage{tensor}
\usepackage{color}
\usepackage{multicol}
\usepackage{bbold}
\usepackage{tikz}
\usetikzlibrary{calc,decorations.markings}

\advance\textheight\topskip
\allowdisplaybreaks[1]

\newcommand\td{\text{d}}

\newcommand{\p}{\partial}

\newcommand{\be}{\begin{equation}}
\newcommand{\ee}{\end{equation}}
\newcommand{\bea}{\begin{eqnarray}}
\newcommand{\eea}{\end{eqnarray}}
\def\nn{\nonumber}
\def\bz{\bar z}

\newcommand*\xbar[1]{%
  \hbox{%
    \vbox{%
      \hrule height 0.5pt 
      \kern0.3ex
      \hbox{%
        \kern-0.0em
        \ensuremath{#1}%
        \kern-0.0em
      }%
    }%
  }%
}
\def\bra#1{\left\langle #1\right|}
\def\ket#1{\left| #1\right\rangle}
\def\>{\rangle} 
\def\<{\langle} 

\DeclareFontFamily{OT1}{rsfs}{} \DeclareFontShape{OT1}{rsfs}{m}{n}{
<-7> rsfs5 <7-10> rsfs7 <10-> rsfs10}{}
\DeclareMathAlphabet{\mycal}{OT1}{rsfs}{m}{n}

\begin{document}
\title{Loop corrections versus marginal deformation in celestial holography}

\author{Song He, Pujian Mao, and Xin-Cheng Mao}

\date{}

\def\mytitle{Loop corrections versus marginal deformation in celestial holography}

\addtolength{\headsep}{4pt}

\begin{centering}


  \textbf{\Large{\mytitle}}

  \vspace{1cm}

  {\large Song He$^{\spadesuit,\heartsuit,\clubsuit,\diamondsuit }$, Pujian Mao$^\P$, and Xin-Cheng Mao$^\clubsuit$ } 

\vspace{0.5cm}

\begin{minipage}{.9\textwidth}\small \it  \begin{center}
     ${}^\spadesuit$ Institute of Fundamental Physics and Quantum Technology, Ningbo University, Ningbo, Zhejiang 315211, China
 \end{center}
\end{minipage}

\vspace{0.3cm}

\begin{minipage}{.9\textwidth}\small \it  \begin{center}
     ${}^\heartsuit$ School of Physical Science and Technology, Ningbo University,\\ Ningbo, 315211, China
 \end{center}
\end{minipage}

\vspace{0.3cm}

\begin{minipage}{.9\textwidth}\small \it  \begin{center}
     ${}^\P$ Center for Joint Quantum Studies and Department of Physics,\\
     School of Science, Tianjin University, 135 Yaguan Road, Tianjin 300350, China
 \end{center}
\end{minipage}

\vspace{0.3cm}

\begin{minipage}{.9\textwidth}\small \it  \begin{center}
    ${}^\clubsuit$ Center for Theoretical Physics and College of Physics,\\ Jilin University, 2699 Qianjin Street,
   Changchun 130012, China
 \end{center}
\end{minipage}

\vspace{0.3cm}

\begin{minipage}{.9\textwidth}\small \it  \begin{center}
     ${}^\diamondsuit$ Max Planck Institute for Gravitational Physics (Albert Einstein Institute),\\
     Am M\"{u}hlenberg 1, 14476 Golm, Germany
 \end{center}
\end{minipage}

\end{centering}

\vspace{0.5cm}

\begin{center}
Emails: hesong@jlu.edu.cn,\, pjmao@tju.edu.cn,\, maoxc1120@mails.jlu.edu.cn
\end{center}

\vspace{0.5cm}

\begin{center}
\begin{minipage}{.9\textwidth}
\textsc{Abstract}: Four-dimensional all-loop amplitudes in QED and gravity exhibit universal Infrared (IR) singularities with a factorization structure. This structure is governed by tree amplitudes and a universal IR-divergent factor representing the exchange of soft particles between external lines. This letter offers a precise dual interpretation of these universal IR-divergent factors within celestial holography. Considering the tree amplitude as the foundation of the celestial conformal field theory (CCFT), these universal factors correspond to marginal deformations with a \textit{double-current} construction in the CCFT. Remarkably, a novel geometric representation of these deformations through topological gauging provides an exact description of transitions within bulk vacuum moduli spaces which extends the celestial holography beyond the perturbative level. Our findings establish a concrete dictionary for celestial holography and offer a holographic lens to understand loop corrections in scattering amplitudes.

\end{minipage}
\end{center}
\thispagestyle{empty}

\section{Introduction}

The holographic principle \cite{tHooft:1993dmi, Susskind:1994vu}, initially formulated within the context of AdS/CFT \cite{Maldacena:1997re}, has been a cornerstone in the quest to understand quantum gravity, proposing that the information within a space can be encoded on its boundary.
Thriving in anti-de Sitter spaces, its extension to flat spacetime has both theoretical and applied necessity urgently \cite{Susskind:1998vk, Polchinski:1999ry,deBoer:2003vf, Arcioni:2003xx, Arcioni:2003td, Solodukhin:2004gs, Barnich:2006av, Guica:2008mu, Barnich:2009se, Barnich:2010eb, Bagchi:2010zz, Bagchi:2012xr}. In flat spacetime, scattering amplitudes are of fundamental importance in perturbative quantum field theory (QFT). They are the main concerns of the recently constructed celestial holography \cite{Pasterski:2016qvg, Pasterski:2017kqt, Pasterski:2017ylz, Raclariu:2021zjz, Pasterski:2021rjz, Pasterski:2021raf, string}, which formally establishes the holographic principle for flat spacetime.

Crucially, Mellin transform establishes a concrete link for celestial holography between 4D massless scattering amplitudes in perturbative QFT and the correlators in CCFT. While the precise properties of the CCFT are yet to be fully defined, which leads to the recent inquiries from various angles, including fundamental CFT characteristics \cite{Schreiber:2017jsr, Fotopoulos:2019tpe, Pate:2019lpp, Arkani-Hamed:2020gyp, Atanasov:2021cje, Crawley:2021ivb, Fiorucci:2023lpb}, symmetries \cite{Fotopoulos:2019vac, Banerjee:2020kaa, Donnay:2020guq, Guevara:2021abz, Himwich:2021dau, Donnay:2021wrk}, loop corrections \cite{Banerjee:2017jeg, Albayrak:2020saa, Gonzalez:2020tpi, Pasterski:2022djr, Donnay:2022hkf, Bhardwaj:2022anh}, double copy relations \cite{Casali:2020vuy, Casali:2020uvr}, supersymmetric extensions \cite{Jiang:2021xzy, Brandhuber:2021nez, Hu:2021lrx}, Carrollian perspectives \cite{Donnay:2022aba, Bagchi:2022emh, Donnay:2022wvx, Bagchi:2023fbj, Saha:2023hsl}, and diverse deformations \cite{Mago:2021wje,Kapec:2022axw, Monteiro:2022lwm, Bu:2022iak, He:2022zcf,Kapec:2022hih,Monteiro:2022xwq, Melton:2022fsf, Fukada:2023ohh}, underscoring the transformative potential of celestial holography.

Notably, current investigations have been predominantly bottom-up by exploring the transformation from amplitude to correlators of CCFT and by examining properties from the dual perspective \cite{Donnay:2023kvm,Pasterski:2023ikd,Donnay:2023mrd}, leaving a top-down quantitative verification of celestial holography, namely exactly matching of independent computations beyond symmetry perspective on both sides, an open challenge.\footnote{A top-down realization \cite{Costello:2022wso,Costello:2022jpg,Costello:2023hmi} of holography for asymptotically flat spacetimes has been recently constructed in the context of twisted holography.} This extremely limits the implications of celestial holography for a broader understanding of scattering amplitudes. Moreover, scattering amplitude is defined by nature perturbatively. Hence, the non-perturbative feature is definitely beyond the current scope of celestial holography.

This letter contributes to the growing body of research in celestial holography by addressing both critical challenges – the quantitative verification of this intriguing framework and its non-perturbative extension. Specifically, we focus on the universal infrared (IR) singularities exhibited by four-dimensional all-loop amplitudes in QED and gravity. The bulk theory's all-loop corrected amplitudes exhibit both ultraviolet (UV) and IR divergences. The IR divergences follow a universal factorization structure~\cite{Weinberg:1965nx} resulting from the exchange of soft virtual gauge bosons and gravitons. The leading IR singularity of the loop amplitude is determined by the tree amplitude and a universal factor.

Considering the tree amplitude as seed CCFT, we systematically establish marginal deformation for the seed CCFT via double-current construction. An exact agreement has been uncovered between the universal part (the leading IR singularity) of the loop QED \cite{Weinberg:1965nx, Arkani-Hamed:2020gyp} and gravity \cite{Naculich:2011ry, Akhoury:2011kq, Bern:2014oka} amplitudes and marginal deformations of the corresponding seed CCFT correlators. Our findings offer a \textit{concrete dictionary} of celestial holography and extend the original avenue of celestial holography. A concrete holographic interpretation of loop corrections to the amplitudes is revealed. This will definitely open a new avenue for the understanding of the structure of loop amplitudes. Then we propose a geometric representation of the marginal deformation, which remarkably equates marginal deformations in the CCFT with bulk vacua's moduli spaces. The celestial holography is thus verified at the non-perturbative level.


\section{4D amplitudes as 2D correlators}

The isomorphism between the 4D Lorentz group and the 2D global conformal group suggests the possibility of restructuring 4D quantum field theory using 2D CFT data. To manifest this connection, the conformal basis was introduced \cite{deBoer:2003vf, Cheung:2016iub, Pasterski:2016qvg, Pasterski:2017kqt}. In the case of masslessness, the transition from momentum space to the conformal basis is achieved through a Mellin transform. Specifically, we parameterize null momenta and polarization vectors in terms of energy $\omega$ and angular parameters $\Vec{z} = (z,\bar{z})$ as follows\footnote{Here we focus on 4D scattering in Minkowski spacetime.
Massless particles enter and exit at past and future null infinity $\mathcal{I}^{\mp}$, parameterized by advanced $v=t+r$ and retarded $u=t-r$ time coordinates. In what follows, we will use stereographic coordinates $(z,\bz)$ to parameterize the celestial plane which is conformally mapped from the celestial sphere.
See, e.g., \cite{He:2019jjk} for the properties of the celestial plane and \cite{Barnich:2016lyg,Compere:2016jwb,Compere:2018ylh,Barnich:2021dta} for the transformations between the celestial plane and sphere.}
\begin{equation} \label{parametrization}
\begin{split}
&q^{\mu} (\omega, \Vec{z}) = \frac{\Tilde{\epsilon} \omega}{2} \Big(1+z\bar{z}, z+\bar{z}, -i(z-\bar{z}), 1-z\bar{z} \Big),\\
&\varepsilon^{+}_{\mu}(q)=\frac{1}{\sqrt{2}}(-\bar{z},1,-i,-\bar{z}),\quad \varepsilon^-_{\mu}(q) =\bar{\varepsilon}^+_{\mu}(q),
\end{split}
\end{equation}
where $\Tilde{\epsilon} = 1, -1$ for outgoing and incoming particles respectively, and $\Vec{z}$ indicates the location on the celestial plane. The $n$-point amplitude of massless particles is identified as a correlator as follows
\begin{align}\label{An}
    \mathcal{A}_n= \bra{out} \mathcal{S} \ket{in} =\<\mathcal{O}_1\cdots \mathcal{O}_n\>,
\end{align}
where $\mathcal{O}_k$ is the annihilation or creation operator of $k$-th particle, which can be recast in terms of operators smeared along null infinity in the extrapolate dictionary \cite{He:2014laa, He:2020ifr, Pasterski:2021dqe}. Mellin transforming the correlator in Eq.~\eqref{An} we have \cite{Pasterski:2017ylz}
\begin{equation} \label{eq:Mellin}
    \<X_n\> = \prod^n_{k=1} \left(\int_0^{\infty} d\omega_k \omega_k^{\Delta_k-1} \right)
    {\cal A}_n,
\end{equation}
where $X_n = \prod^n_{k = 1} \mathcal{O}_{\Delta_k,s_k}$

Perturbatively, the amplitude in momentum space can be expressed as a series in powers of the bulk theory's coupling constant $g$: $\mathcal{A}_n = \sum_{\ell = 0}^{\infty} \mathcal{A}_n^{(\ell)}$, where
\begin{equation} \label{generic}
\mathcal{A}_n^{(\ell)} = \frac{g^{\ell}}{\ell!} \left[  \frac{(B_n)^{\ell}}{\epsilon^{\ell}} \mathcal{A}_n^{(0)}  + \mathcal{N}^{(\ell)} \right].
\end{equation}
Here, $\ell$ denotes the loop order, and $\epsilon$ acts as the dimensional regularization parameter governing IR singularities. The right-hand side of Eq. \eqref{generic} consists of two parts: the first one, primarily universal, arises from Abelian Feynman diagrams \cite{Arkani-Hamed:2020gyp, Naculich:2011ry}. The second term encompasses non-universal infrared contributions, which are power-suppressed in the dimensional regularization.

For scalar QED, $g_e = - \frac{e^2}{8 \pi^3}$, and the leading IR factor \cite{Weinberg:1965nx, Kalyanapuram:2021bvf} from Eq. \eqref{generic} is
\begin{equation} \label{eq:LoopQED}
    \begin{split}
        B_n^e & = \pi \sum_{i, j} Q_i Q_j \log \frac{-2q_i \cdot q_j}{\mu^2} \\
        & = \pi \sum_{i,j} Q_i Q_j \log |z_{ij}|^2
    \end{split}
\end{equation}
where $Q_k$ represents the electric charge of the $k$-th particle, $z_{ij} = z_i - z_j$, and $\mu$ is an arbitrary energy scale. The terms involving the energy scale can be eliminated by utilizing the conservation law $\sum_k Q_k = 0$, with the parameterization provided in the second equality.
Implementing Mellin transform to the full amplitude in Eq. \eqref{generic} in QED, denoted as $\mathcal{A}_n^e$, one obtains\footnote{The parametrization of the factor in the exponential here is different than the one in \cite{Arkani-Hamed:2020gyp} because we keep the collinear-divergent terms, see, e.g., the comments in footnote 11 of that reference. The aim of the present work is to reveal a concrete correspondence between 4D amplitudes and 2D correlators. Subtractions in 4D and 2D are different in general, and we will not consider the collinear divergences in this paper.}
\begin{equation} \label{qed}
    \<X_n\>^{(\ell)}_e = \frac{g_e^{\ell}}{\ell!} \left[ \frac{(\sigma_n^e)^{\ell}}{\epsilon^{\ell}} \<X_n\>^{(0)}_e + {N}_e^{(\ell)} \right],
\end{equation}
where $\<X_n\>_e^{(0)}$ is the correlator of CCFT that is Mellin transform of the tree QED amplitude, $\sigma_n^e = B_n^e$, and ${N}_e^{(\ell)}$ corresponds to Mellin transformed non-universal contributions $\mathcal{N}_e^{(\ell)}$.

For gravity, the coupling constant is set to $g_G = \frac{G_N}{16 \pi^2}$. The leading universal IR factor from Eq. \eqref{generic} is \cite{Weinberg:1965nx,Naculich:2011ry,Akhoury:2011kq,Bern:2014oka}
\begin{equation} \label{eq:allloop}
    \begin{split}
        B_n^G & = 8 \pi \sum_{i, j} q_i \cdot q_j \log \frac{-2q_i \cdot q_j}{\mu^2} \\
        & = - 4 \pi \sum_{i, j} |z_{ij}|^2 \omega_i \omega_j \log |z_{ij}|^2.
    \end{split}
\end{equation}
Implementing Mellin transform to the full graviton amplitude $\mathcal{A}^G_n$ yields
\begin{equation} \label{gravity}
    \begin{split}
        & \<X_n\>^{(\ell)}_G = \frac{g_G^{\ell}}{\ell!} \left[ \frac{(\sigma_n^G)^{\ell}}{\epsilon^{\ell}} \<X_n\>^{(0)}_G + {N}_G^{(\ell)} \right], \\
        & \sigma_n^G = \pi \sum_{i, j} |z_{ij}|^2 \mathcal{G}_i \mathcal{G}_j \log |z_{ij}|^2
    \end{split}
\end{equation}
where
\begin{equation}\label{G}
    \mathcal{G}_k \mathcal{O}_{\Delta_k, s_k} = \mathcal{O}_{\Delta_k + 1, s_k},
\end{equation}
and ${N}_G^{(\ell)}$ corresponds to Mellin transformed $\mathcal{N}_G^{(\ell)}$ for gravity amplitude.


\section{Seed theories and Currents} \label{sec-QED}

We select for our seed CCFT correlator the Mellin transform of the tree-level scalar QED amplitude $\<X_n\>^{(0)}_e$ and gravity amplitude $\<X_n\>^{(0)}_G$, respectively. The seed CCFTs admit symmetry currents originated from soft theorems. In the conformal basis, they are represented by conformally soft operator $\mathcal{O}_{\Delta,s}$ with an integer conformal dimension $\Delta$. The conformally soft operators of relevance for constructing deformations in the present work are with dimension 1. The Ward identity of such soft operators $\mathcal{O}_{1,s}$ are divergent, which can be regulated by slightly shifting the conformal dimension similar to dimensional regularization as $\Delta = 1 + \epsilon$ \cite{Adamo:2019ipt},
\begin{equation} \label{eq:Ward}
    \<\mathcal{O}_{1, s} X_n\>^{(0)} = 
    \frac{W^{(0) \pm}}{\epsilon} \<X_n\>^{(0)} + O (\epsilon^0),
\end{equation}
where the leading factor $W^{(0) \pm}$ is universal with non-universal subleading corrections from the regularization.

{In scalar QED, the seed CCFT features $U(1)$ symmetry \cite{He:2015zea, Kapec:2017gsg, Kapec:2021eug} is derived from $\mathcal{O}^e_{1, \pm 1}$ with the leading universal factors
\begin{equation} \label{srelation}
    W^{(0) +}_e = \sum_{k = 1}^n \frac{Q_k}{z - z_k}, \quad W^{(0) -}_e = \sum_{k = 1}^n \frac{Q_k}{\Bar{z} - \Bar{z}_k}
\end{equation}
which corresponds to the leading soft photon theorem originally obtained in \cite{Low:1954kd, Gell-Mann:1954wra, Low:1958sn, Weinberg:1965nx}.
The soft current in the context of 2D CCFT \cite{Kapec:2017gsg, Kapec:2021eug} is $(J, \Bar{J}) = (\mathcal{O}^e_{1, +1}, \mathcal{O}^e_{1, -1})$ with the Ward identity in Eq. \eqref{eq:Ward},
where $\mathcal{O}^e_{1, -1}$ is the shadow \cite{Nande:2017dba,Kapec:2021eug} of $\mathcal{O}^e_{1, +1}$, emerging another copy of $U(1)$ symmetry. Therefore, the symmetries generated by $\mathcal{O}^e_{1, \pm 1}$
can be split into a $U(1) \times U(1)$ symmetry, where $J^1_A = (J,0)$ and $J^2_A = (0, \Bar{J})$. The indices $1$ and $2$ distinguish the currents from different copies of $U(1)$.
It is worth noting that the operator product expansion (OPE) of $JJ$ and $J\Bar{J}$ vanishes at the leading order of the regularization \cite{Pate:2019lpp, Fotopoulos:2019vac}. This observation can also be understood more intuitively: the leading soft photon theorem originates from large gauge transformations, and the symmetry algebra is Abelian with no central extension \cite{Strominger:2013lka, He:2014cra}.
From the boundary perspective, these currents arise from the fluctuations of the gauge field $A^i_{A}$ residing on the celestial sphere, given by $J^A_i = \frac{\delta S_e}{\delta A^i_A}$, with $i = 1,2$. Here, $S_e$ represents the CCFT action incorporating $U(1) \times U(1)$ symmetry.}

When a soft graviton is emitted in a scattering process, the amplitude satisfies the Weinberg's soft theorem \cite{Weinberg:1965nx}.
Correspondingly, one can introduce conformal soft operators $\mathcal{O}^G_{1,\pm2}$ with dimension 1 with the insertion in 2D correlators as
\begin{equation} \label{eq:WG}
    W^{(0)+}_G = \sum^n_{k=1} \frac{\bar{z}-\bar{z}_k}{z-z_k} \mathcal{G}_k, \quad W^{(0)-}_G = \sum^n_{k=1} \frac{z-z_k}{\bar{z}-\bar{z}_k} \mathcal{G}_k,
\end{equation}
where $\mathcal{G}_k$ is defined in Eq. \eqref{G}.
Then we define the soft current as $P = \mathcal{O}^G_{1, +2}$, and its shadow \cite{Kapec:2021eug}, $\Bar{P} = \mathcal{O}^G_{1, -2}$, which are components of a traceless current $P_{AB}$. It is important to note that $P_{AB}$ is Abelian in the leading order because of the vanishing of the $PP$ and $P\Bar{P}$ OPEs \cite{Pate:2019lpp, Fotopoulos:2019vac}.
However, it is not a Kac-Moody current, as discussed in \cite{Kapec:2017gsg}.
The scattering of soft gravitons from bulk to boundary induces fluctuations in the boundary metric $\gamma_{AB}$, associated with different soft graviton modes, each dual to distinct currents. These fluctuations are represented in the basis $E^a_{\ A}$, enabling the decomposition of the 2D transverse metric as $\gamma_{AB} = g_{ab} E^a_{\ A} E^b_{\ B}$, where $a$ and $b$ indices use $g_{ab}$ for raising and lowering.
We denote $E^a_{\ A}$ as the basis for the soft graviton mode.
$P^a_{\ A}$ is defined as
$P^a_{\ A} = P^B_{\ A} E^a_{\ B} = \frac{\delta S_G}{\delta E_a^{\ A}}$, where $S_G$ represents the action of the CCFT with symmetries generated by $P_{AB}$, related to bulk supertranslations.

\section{Deformations}

We propose that a standard 2D CCFT double-trace deformation captures the universal IR structure of the all-loop corrected amplitude in Eq. \eqref{generic}. This deformation, constructed by two commutative currents, can be defined as \cite{Cardy:2019qao}
\begin{equation} \label{eq:genericS}
    \frac{\p S^{[\lambda]}}{\p \lambda} = \int \td^2 x \ \mathcal{O}_{j\Bar{j}}^{[\lambda]}, \quad \mathcal{O}_{j\Bar{j}}^{[\lambda]} = \frac{1}{2} \epsilon_{\Bar{a} \Bar{b}} \epsilon^{AB} j^{\Bar{a}[\lambda]}_{A} j^{\Bar{b} [\lambda]}_{B}
\end{equation}
Here, $S^{[\lambda]}$ serves as the action deformed from the seed CCFT,\footnote{It is important to note that calculating the universal part of deformed correlators only requires the data of correlators in the seed CCFT, the precise form of the seed CCFT action is not needed.} $\lambda$ is a coupling constant, $\Bar{a}$ is an abstract index, which can be chosen as $i$ or $a$ in this letter, $\epsilon_{AB}, \epsilon^{ab}$ are Levi-Civita tensors while $\epsilon_{ij}$ is Levi-Civita symbol.

In scalar QED, we utilize the soft current $J^i_A$ to induce a dimensionless marginal deformation within the CCFT.
{The $J \Bar{J}$ deformed action by specifying the abstract indices as $\Bar{a} \Bar{b} = ij$ and the current as $j^{\Bar{a}}_A = J^i_A$ within the generic form Eq. \eqref{eq:genericS}.\footnote{If one considers the soft currents as classical fields, the deformation of the action is precisely the inverse shadow of the effective action of the soft photon modes in Ref. \cite{Kapec:2021eug}. There the shadow transformation is introduced by the propagator in the effective action. However, in our construction, a shadow transformation is not involved. The situation is similar to the gravity case in the next section.}
Applying the Ward identities in Eq.
\eqref{eq:Ward} and Eq. \eqref{srelation}, one obtains the $\ell$-th order correction of the deformed correlator
\be
     \<X_n\>^{(\ell)}_{[\lambda_e]}
     = \frac{1}{\ell!} \left( - \frac{\lambda_e}{\epsilon} \right)^{\ell} \bigg[ \bigg( - \frac{1}{\epsilon} \sum_{i,j = 1}^n Q_i Q_j I_{i\Bar{j}}^{1 \Bar{1}} \bigg)^{\ell} \< X_n\>^{(0)}_e + {N}^{(\ell)}_{\lambda_e} \bigg]
\ee
where ${N}^{(\ell)}_{\lambda_e}$ is the non-universal part which is from the flow effect of the current $J_A^i$ and is power-suppressed in the expansion of $\epsilon$.} The integral $I_{i\Bar{j}}^{r \Bar{s}}$ is defined as
\begin{equation} \label{integral}
    I_{i \bar{j}}^{r \bar{s}} = \frac{i}{2} \int \frac{d^2z}{(z-z_i)^r (\bar{z}-\bar{z}_j)^s},
\end{equation}
which can be computed using the Stokes formula, as detailed in \cite{He:2022zcf}, and their evaluations are provided in the Appendix \ref{A}.
Finally, 
\begin{equation} \label{qedcorrelator}
    \<X_n\>^{(\ell)}_{[\lambda_e]} = \frac{1}{\ell!} \left( - \frac{\lambda_e}{\epsilon} \right)^{\ell} \left[ \frac{(  \sigma_n^e )^{\ell} }{\epsilon^{\ell}} \< X_n\>^{(0)}_e + {N}^{(\ell)}_{\lambda_e} \right],
\end{equation}
where charge conservation lets us eliminate any scale for the celestial plane introduced in the evaluation of Eq. \eqref{integral}.
Here, $\sigma_n^e$ 
is precisely the one introduced in Eq. \eqref{qed}.
Hence, by identifying the regularization parameters in Eq. \eqref{generic} and Eq. \eqref{eq:Ward} and setting the deformation coupling constant
$ \lambda_e$ as $- \epsilon g_e$, an exact equivalence emerges between the universal IR divergent factor of the all-loop amplitude and the $J \Bar{J}$ deformed correlator.

A parallel correspondence holds true for gravity. Utilizing the operator $P_{AB}$, we formulate a 2D marginal deformation, addressing the 4D loop correction of amplitudes exhibiting an infrared leading divergence. The action governing the double-trace $P \Bar{P}$ deformation specifies the abstract indices as $\Bar{a} \Bar{b} = ab$, and designates the current as $j = P$ in Eq. \eqref{eq:genericS}, with $\lambda_G$ representing a dimensionless coupling constant. The $\ell$-th order correction of the $P \Bar{P}$ deformed correlator can be obtained by configuring $j \Bar{j} = P \Bar{P}$.
Employing the Ward identities in Eq.
\eqref{eq:Ward}, Eq. \eqref{eq:WG} and evaluating the integrals $I_{i\Bar{j}}^{r \Bar{s}}$ (detailed in the Appendix \ref{A}), yields the following result:
\begin{equation} \label{gravitycorrelator}
    \<X_n\>^{(\ell)}_{[\lambda_G]} = \frac{1}{\ell!} \left( - \frac{\lambda_G}{\epsilon} \right)^{\ell} \left[ \frac{(  \sigma_n^G )^{\ell}}{\epsilon^{\ell}} \< X_n\>^{(0)}_G + {N}^{(\ell)}_{\lambda_G} \right],
\end{equation}
where $\sigma_n^G$ is the same as that introduced in Eq. \eqref{gravity}. By identifying $\lambda_G$ as $- \epsilon g_G$,\footnote{In the main text, we have assumed that the length scale of the celestial plane is $1$, using stereographic coordinates on the Riemann sphere $(z,\bar{z})$. One could also restore the length scale $r_0$ to regulate the collinear limits, in which case the identification should be
$\lambda_G = - \frac{\epsilon G_N}{16 \pi^2 r_0^2}$ for the identifications in Eq. \eqref{gravitycorrelator}.}
the $\ell$-loop corrected 4D gravity amplitudes, particularly for the leading term of IR divergence, can be regarded as correlation functions in a seed CCFT with the $\ell$-th marginal deformation ${P \Bar{P}}$. For consistency, we reformulate soft theorems \cite{Donnay:2018neh, Adamo:2019ipt, Puhm:2019zbl, Guevara:2019ypd} within the deformed CCFT. It is shown that the leading soft theorem does not flow under the deformation. This consolidates the fact that the leading IR singularity in Eq. \eqref{generic} is universal. Our rigorous confirmation also shows that the deformed Ward identity of the stress tensor, arising as a shadow transformation of the deformed subleading soft theorem, precisely agrees with the amplitude-based derivation \cite{He:2017fsb}. Explicit computations are available in the Appendix \ref{B}-\ref{E}.


\section{Geometric realization of Moduli spaces of bulk vacua}

In standard holographic scenarios, marginal deformations in the boundary conformal field theory correspond to continuous moduli spaces of bulk gravitational vacua. Recent works \cite{Kapec:2022axw,Kapec:2022hih} in celestial holography has extended this idea to transitions in bulk vacua at the perturbative level, characterized by introducing coherent states of zero-energy soft particles. The marginal deformation discussed in the previous sections aligns with such transitions, representing changes in electromagnetic or gravitational field vacua. To explore these transitions in more detail from a non-perturbative perspective, we employ a geometric realization of the marginal deformation, as proposed by \cite{Dubovsky:2017cnj, Cardy:2018sdv, Dubovsky:2018bmo, Dubovsky:2023lza}.

In celestial holography, the nature of the symmetry breaking and relation of the degenerate vacua are very different from the standard case of spontaneous breaking of a global symmetry \cite{Strominger:2017zoo}. The parameter that generates the spontaneously broken symmetry
is an arbitrary local function on the celestial sphere. The vacuum
state is changed by soft particle creation, which is classically measured by the memory effect.
At the classical level, the degenerate vacua all have vanishing field strength tensor $F_{\mu\nu}$ or curvature tensor $R_{\mu\nu\alpha\beta}$, and are characterized by asymptotic symmetry transformation, see, e.g., detailed analysis in Chapter 2.11 of \cite{Strominger:2017zoo}.

Suppose that the action $S [\phi, F]$ is associated with the complete seed CFT, where $\phi$ denotes a collection of arbitrary fundamental fields and $F$ is the background gauge field. A double current  deformation can be viewed as a form of topological gauging \cite{Dubovsky:2017cnj, Cardy:2018sdv, Dubovsky:2018bmo, Dubovsky:2023lza},
\begin{equation} \label{eq:massive}
    \begin{split}
        S^{[\lambda]} & = S [\phi, F] - \frac{1}{2 \lambda} S_M[F, Y], \\
        S_M & = \int \td^2 x \ \epsilon^{AB} \epsilon_{\Bar{a} \Bar{b}} \left( F^{\Bar{a}}_{ A} - Y^{\Bar{a}}_{\ A} \right) \left( F^{\Bar{b}}_{ B} - Y^{\Bar{b}}_{\ B} \right),
    \end{split}
\end{equation}
where $F^{\Bar{a}}_{ A}$ now is promoted into a dynamical gauge field, and
$Y^{\Bar{a}}_{\ A}$ is the auxiliary field. Correspondingly, the double current deformed action results from integrating over $F^{\Bar{a}}_{ A}$ using a leading semi-classical approximation with the saddle point condition \cite{Dubovsky:2023lza}. We set $Y^{\Bar{a}}_{\ A}$ to represent the reference background gauge field, ensuring the proper recovery of the undeformed theory as $\lambda$ approaches zero. The path integral transformation of Eq. \eqref{eq:massive} provides a comprehensive quantum definition for these deformations \cite{Dubovsky:2023lza}. Different saddle points in the gauge field $F^{\Bar{a}}_{ A}$ lead to distinct action decay patterns, distinguishing theories with unique vacua. Notably, specific saddle point choices automatically induce $J \Bar{J}$ and $P \Bar{P}$ deformations
\begin{equation} \label{eq:B*}
    F^{\Bar{a}}_{* A} = Y^{\Bar{a}}_A + \lambda \epsilon^{\Bar{a} \Bar{b}} \epsilon_{AB} j^{B [\lambda]}_{\Bar{b}}.
\end{equation}
In a holographic understanding, the $F_*$ can be regarded as the source field coming from the bulk, which generates fluctuation to the vacua of the bulk.

For scalar QED, $Y^{\Bar{a}}_{\ A}$ is the background (undeformed) gauge field $A^{i}_{\ A}$, so the $J \Bar{J}$ deformation will change the gauge field in the bulk from $A^i_A$ to $F^i_{* A}$
\begin{equation}
    A^i \rightarrow F^i_* (\lambda_e) = A^i + \td \chi^i (\lambda_e),
\end{equation}
where $\lambda_e \epsilon^{ij} \epsilon_{AB} J^{B [\lambda_e]}_j = \td \chi^i_A$ since $\td (\lambda_e \epsilon^{ij} \epsilon_{AB} J^{B [\lambda_e]}_j) = 0$ for conserved currents. This indicates that the $J \Bar{J}$ deformation on the boundary is equivalent to implementing a large gauge transformation in bulk with dynamic terms, which leads to the transition between inequivalent vacua in bulk \cite{He:2014cra, Strominger:2017zoo}.

For gravity, $Y^{\Bar{a}}_{\ A}$ is the background basis $E^a_{\ A}$ of the undeformed metric, i.e., the Minkowskian bulk metric. It will be formally changed through the $P\Bar{P}$ flow. In particular, the transverse metric will be switched from $\gamma_{AB}$ to
\begin{equation}
\begin{split}
    \gamma_{AB}^{[\lambda_G]} &= g_{ab} F^a_{* A} F^b_{* B} \\
    &= \gamma_{AB} + 2 \lambda_G \hat{P}_{AB}^{[\lambda_G]} + \lambda_G^2  \hat{P}_{AC}^{[\lambda_G]} \hat{P}^{[\lambda_G]}_{BD} \gamma^{CD},
    \end{split}
\end{equation}
where $\hat{P}_{AB}^{[\lambda_G]} = P_{AB}^{[\lambda_G]} - \gamma_{AB} P^{C [\lambda_G]}_{\ C}$. As a marginal deformation, the current $\hat{P}_{AB}^{[\lambda_G]}$ should obey the same properties as the seed theory from a non-perturbative perspective, which leads to the following decomposition that the current must satisfy,
\be
\hat{P}_{AB}^{[\lambda_G]} = - 2 \p_A \p_B C(z,\bz) + \gamma_{AB} \p^D \p_D C(z,\bz).
\ee
Remarkably, the supertranslation shifts the transverse metric as \cite{Compere:2016jwb,Compere:2018ylh}
\begin{equation}
    \gamma_{AB}^{\text{vac}} = \gamma_{AB} + \frac{1}{\rho} C^{\text{vac}}_{AB} + \frac{1}{4 \rho^2} C_{AC}^{\text{vac}} C_{BD}^{\text{vac}} \gamma^{CD},
\end{equation}
where the traceless vacuum shear tensor $C^{\text{vac}}_{AB}$ satisfies the same type of decomposition as the current $\hat{P}_{AB}^{[\lambda_G]}$. After balancing the dimension, we can identify $\hat{P}_{AB}^{[\lambda_G]}$ as $\frac{C_{AB}^{\text{vac}}}{2 G_N \epsilon}$ in the bulk. Together with the identification $\lambda_G = \frac{\epsilon G_N}{\rho}$, one obtains
 \begin{equation} \label{eq:gammaidenti}
    \gamma_{AB}^{\text{vac}} = \gamma_{AB}^{[\lambda_G]}.
\end{equation}
This verifies that the $P \Bar{P}$ deformation on the boundary is equivalent to implementing a supertranslation in bulk, a dynamical coordinates transform. The vacuum of standard Minkowskian spacetime will flow to the inequivalent vacua \cite{Compere:2016jwb, Strominger:2017zoo,Compere:2018ylh}.

Finally, it's worth noting that supertranslations and large gauge transformations in the bulk account for the exchange of virtual soft gravitons and gauge bosons, respectively \cite{Weinberg:1965nx, Strominger:2017zoo, Kalyanapuram:2021bvf}. This exchange generates the leading IR divergence in the amplitudes for scalar QED and gravity \cite{Nande:2017dba, Himwich:2020rro, Kapec:2021eug}. Therefore, it's reasonable to infer that the IR-divergent component of the amplitude relation in Eq. \eqref{generic} corresponds to the deformed action on the boundary in Eq. \eqref{eq:massive} near the saddle point $F_*$, confirming the consistency of our framework with previous research, especially \cite{Zamolodchikov:1986gt, Dixon:1987bg, Seiberg:1988pf, Kapec:2022axw,Kapec:2022hih}.

\section{Concluding remarks}

This paper establishes a precise correspondence between the leading-order universal infrared divergences in 4D massless scattering amplitudes within the bulk theory and marginal deformations of correlators in 2D CCFT.
In the context of QED and gravity, we determine the soft current, $J$ or $P$, by invoking the leading soft photon theorem \cite{Kapec:2017gsg, He:2015zea, Nande:2017dba, Kapec:2021eug} or soft graviton theorem \cite{Kapec:2017gsg, Donnay:2018neh, Himwich:2020rro}, respectively. 
By deforming the seed CCFT with the composite operator $J\Bar{J}$ or $P\Bar{P}$, we uncover a universal part of the deformed correlators expressed in a factorial form. Notably, there is an exact agreement between the correlator relations and the Mellin transform of the universal part of the amplitude relations in Eq. \eqref{generic}, underscoring the remarkable fact that the universal part of the loop amplitude can be precisely reproduced by the tree amplitude and a marginal deformation in the context of celestial holography. Lastly, we propose a geometric representation of the marginal deformation, linking it with the moduli spaces of bulk vacua, thereby offering a promising avenue to understand the interplay between boundary CCFT and bulk theories.

In summary, we offer quantitative verification for celestial holography at perturbative and non-perturbative levels, opening avenues for future exploration.

Firstly, within our framework, soft theorems, known to extend beyond the leading order \cite{Low:1954kd,Low:1958sn,Gross:1968in,White:2011yy,Cachazo:2014fwa}, allow systematic construction of marginal deformations using subleading soft currents. The Mellin transformation of resulting CCFT correlator relations is poised to unveil insightful holographic amplitude relations, potentially discovering novel results for loop amplitude. Despite amplitude relations being expressed in the low-energy expansion form in Eq. \eqref{generic}, they closely relate to the classical limit of loop amplitudes, crucial for recent amplitude-based gravitational wave investigations \cite{Cheung:2018wkq,Bern:2019nnu,Bern:2019crd,Cheung:2020gyp,Bern:2020buy,Cheung:2020sdj,Herrmann:2021lqe,Bern:2021dqo,Bern:2021yeh,Bern:2022kto,Mougiakakos:2022sic,Aoude:2022thd,FebresCordero:2022jts,Kosower:2018adc,Cristofoli:2021vyo,Adamo:2022qci}.

Secondly, utilizing the geometric representation to identify moduli spaces of bulk vacua corresponding to subleading soft theorems' marginal deformations is crucial. This uniquely determines the symmetry transformation of the bulk solution phase space, addressing distinct proposals in the literature \cite{Lysov:2014csa,Campiglia:2016hvg,Conde:2016csj,Kapec:2014opa,Campiglia:2014yka,Campiglia:2015yka,Campiglia:2016jdj,Campiglia:2016efb,Conde:2016rom}.

\section*{Acknowledgments}
The authors thank Yangrui Hu, Wenliang Li, Zhengwen Liu, Huajia Wang, Jun-Bao Wu, Ellis Ye Yuan, and Hongbao Zhang for their valuable discussions and comments. The authors would like to express their sincere gratitude to Sabrina Pasterski for her invaluable assistance throughout this work, which includes enlightening private communications, as well as valuable comments and suggestions on the draft. We also thank the anonymous reviewer for drawing our attention to the moduli spaces of vacua in the corresponding dual bulk theory. This work is partly supported by the National Natural Science Foundation of China under Grant No.~12075101, No.~12235016, No.~11935009 and No.~11905156. S.H. is grateful for financial support from the Fundamental Research Funds for the Central Universities and the Max Planck Partner Group.


\appendix

\section{Evaluation of the integrals in the deformed correlators}
\label{A}

To obtain the deformed correlators in the main text, we need to compute the following integrals \begin{equation}
    I_{i \bar{j}}^{r \bar{s}} = \frac{i}{2} \int \frac{d^2z}{(z-z_i)^r (\bar{z}-\bar{z}_j)^s}.
\end{equation}
Here we will follow the strategy for computing surface integral in stereographic coordinates detailed in \cite{He:2022zcf}. Analytic continuations of the integration over $\mathbb{CP}^1$ to a double contour integral are also discussed in~\cite{Callebaut:2019omt,He:2020udl,Donnay:2021wrk,Banerjee:2022wht}.
 The key step for our prescription is the application of the Stokes' theorem. A contour integral along the boundaries of the surface can be written as a surface integral
\begin{equation}\label{eq:int dz P}
\oint dz P=-\int d^2z \p_{\bar{z}}P, \qquad  \oint d\bar{z}P=\int d^2z \p_z P .
\end{equation}
Then we can rewrite the surface integral as
\begin{equation}\label{surface}
\int d^2z=\oint d\bar{z} \int dz=-\oint dz \int d\bar{z},
\end{equation}
where one should consider the second integral $\int \td z$ or $\int \td \bz$ as indefinite integral.
The integrand for the first contour integral is still a function of both $z$ and $\bz$. But once the second indefinite integral is done, the first integral is a contour integral on the boundaries where $z$ and $\bz$ are related. One can simply replace $z$($\bz$) by $\bz$($z$) from the constraint of the boundary.

\begin{figure}[ht]
    \begin{center}
        \begin{tikzpicture}
\draw (2,2) arc(5:356:2);\draw (0.5,2) arc(18:342:0.5);\draw (0.5,3) circle(0.5);
\fill (0.5,3) circle (2pt);\fill (0.07,1.85) circle (2pt);
\node[above] at (0.5,3) {$z_{mn}$};\node[below] at (0,1.9) {$0$};\node[above] at (1.5,0) {$\Lambda$};\node[below] at (0.5,1.5) {$\epsilon_0$};\node[left] at (0,3) {$\epsilon_1$}; \node[above] at (1.25,2) {$l_1$}; \node[below] at (1.25,1.7) {$l_2$};
\draw [-latex] (0,-0.185)--(0.0001,-0.185);\draw [-latex] (0.0001,3.82)--(0,3.82);\draw [-latex] (1,3.0001)--(1,2.9999);\draw [-latex] (-0.48,1.88)--(-0.48,1.8801);
\draw (0.5,2)--(2,2);\draw (0.5,1.7)--(2,1.7);
\draw [-latex] (1,2)--(1.25,2);\draw [-latex] (1.5,1.7)--(1.25,1.7);
\end{tikzpicture}
\caption{The contour of the $I_{m \bar{n}}^{1 \bar{1}}$ integral.} \label{fig:Contour}
    \end{center}
\end{figure}
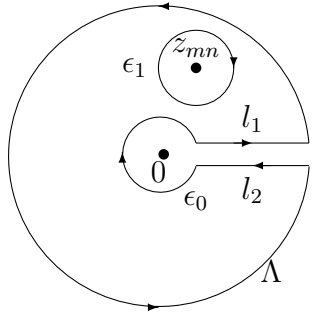

\noindent{\bf Evaluation of $r=s=1$ :}
To evaluate the integral, we shift the parameters $z,\bz$ and use Stokes' formula
\begin{equation}
\begin{split}
    I_{i \bar{j}}^{1 \bar{1}}=&\frac{i}{2}\int\frac{d^2z}{(z-z_i) (\bar{z}-\bar{z}_j)}=\frac{i}{2}\int\frac{d^2z}{(z-z_{ij}) \bar{z}}\\
    =&-\frac{i}{2}\oint dz \int\frac{d\bar{z}}{(z-z_{ij}) \bar{z}}=-\frac{i}{2}\oint dz \frac{\log \bar{z}}{z-z_{ij}}.
\end{split}\end{equation}
The integrand has a branch cut between $z=0\rightarrow\infty$ and a pole at $z=z_{ij}$ (see fig. \ref{fig:Contour} for the integral contour).
Therefore, we divide $I_{i \bar{j}}^{1 \bar{1}}$ into 4 parts
\begin{equation}
    I_{i \bar{j}}^{1 \bar{1}}=-\frac{i}{2}\oint dz \frac{\log \bar{z}}{z-z_{ij}} = I_{\Lambda} + I_{l_1+l_2} + I_{\epsilon_0} + I_{\epsilon_1},
\end{equation}
where
\begin{equation}
    \begin{split}
        I_{\Lambda}= & \frac{1}{2}\int^{2\pi}_0 \Lambda e^{i\theta} d\theta \frac{\log\Lambda-i\theta}{\Lambda e^{i\theta}-z_{ij}} = \frac{1}{2}\int^{2\pi}_0 d\theta (\log\Lambda-i\theta)+\frac{z_{ij}}{2} \int^{2\pi}_0 d\theta \frac{\log\Lambda-i\theta}{\Lambda e^{i\theta}-z_{ij}}\\
        =&\pi \log \Lambda-\pi^2 i+0;
    \end{split}
\end{equation}
\begin{equation}
    I_{\epsilon_0}=-\frac{1}{2}\int^{2\pi}_0 \epsilon_0 e^{i\theta} d\theta \frac{\log\epsilon_0-i\theta}{\epsilon_0 e^{i\theta}-z_{ij}}\rightarrow 0;
\end{equation}
\begin{equation}
    \begin{split}
        I_{\epsilon_1}=&\frac{i}{2}\oint_{|z-z_{ij}|=\epsilon_1} dz\frac{\log \left(\bar{z}_{ij}+\frac{\epsilon_1^2}{z-z_{ij}}\right)}{z-z_{ij}} =\frac{i}{2}\oint_{|z'|=\epsilon_1} \frac{dz'}{z'}\log \left(\bar{z}_{ij}+\frac{\epsilon_1^2}{z'}\right)\\
        =&\frac{i}{2}\oint_{|z'|=\epsilon_1} \frac{dz'}{z'}\left[\log \bar{z}_{ij}+\log \left(1+\frac{\epsilon_1^2}{z'\bar{z}_{ij}}\right)\right]= -\pi \log \bar{z}_{ij}-\frac{i}{2} \sum_{n=1}^{\infty} \frac{1}{n} \oint_0 \frac{dz}{z^{n+1}}\left(\frac{\epsilon_1^2}{\bar{z}_{ji}}\right)^n\\
        =&-\pi \log \bar{z}_{ij}-0;
    \end{split}
\end{equation}
\begin{equation}
    \begin{split}
        I_{l_1+l_2}=&-\frac{i}{2}\int^{\Lambda}_{\epsilon_0} dx\frac{\log x}{x-z_{ij}}+\frac{i}{2} \int^{\Lambda}_{\epsilon_0} dx\frac{\log x-2 \pi i}{x-z_{ij}}\\
         =& \pi \int^{\Lambda}_{\epsilon_0} \frac{dx}{x-z_{ij}}=\pi \log \Lambda-\pi \log z_{ji}.
    \end{split}
\end{equation}
Summing these contributions, we obtain
\begin{equation} \label{eq:I11}
    \begin{split}
        I_{i \bar{j}}^{1 \bar{1}}= & 2\pi \log \Lambda-\pi \log\left(-|z_{ij}|^2\right)-\pi^2 i=2 \pi \log \Lambda-\pi\left[ \log\left(-|z_{ij}|^2\right)-\pi i\right]\\
        =&2 \pi \log \Lambda - \pi \log \left(-|z_{ij}|^2e^{i\pi}\right) = - \pi \log \left( \frac{|z_{ij}|^2}{\Lambda^2} \right).
    \end{split}
\end{equation}
The same integral is evaluated by the dimensional regularization scheme in \cite{He:2019vzf}, where the result is
\begin{equation}
    I_{i \bar{j}}^{1 \bar{1}}\big|_{d=2+\widetilde{\epsilon}}=2\pi \left(\frac{-1}{\widetilde{\epsilon}}-\log |z_{ij}|-\frac{\log \pi +\gamma}{2}\right),\quad \widetilde{\epsilon}\rightarrow-0,
\end{equation}
where $\gamma$ is Euler constant. To connect to our result, one just needs to set
\begin{equation}
    -\frac{1}{\widetilde{\epsilon}}-\frac{\log \pi +\gamma}{2}=\log \Lambda ~~ \rightarrow  ~~ +\infty.
\end{equation}

\noindent{\bf Evaluation of $r=1,\ s = 0$ :}
Shifting the integration parameter, we find
\begin{equation}\label{integralresult}
    I_{m \bar{n}}^{1 \bar{0}} = \frac{i}{2} \int \frac{d^2z}{z-z_m}= \frac{i}{2} \int \frac{d^2z}{z} = I_{\Lambda} - I_{\epsilon},
\end{equation}
where the contour is illustrated in fig \ref{fig:I10}, and
\begin{equation}
    I_{\Lambda} = - \frac{i}{2} \oint_{\Lambda} dz \frac{\Bar{z}}{z} = - \frac{i}{2} \Lambda^2 \oint_{z=0} \frac{dz}{z^2} = 0, \quad I_{\epsilon} = - \frac{i}{2} \oint_{\epsilon} dz \frac{\Bar{z}}{z} = - \frac{i}{2} \epsilon^2 \oint_{z=0} \frac{dz}{z^2} = 0.
\end{equation}
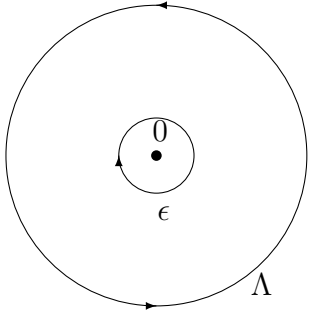
\begin{figure}[ht]
    \begin{center}
        \begin{tikzpicture}
\draw (2,2) circle (2);\draw (2,2) circle (0.5);
\fill (2,2) circle (2pt);
\node[above] at (2.05,2.05) {$0$};\node[above] at (3.4,0) {$\Lambda$};\node[above] at (2.1,1) {$\epsilon$};
\draw [-latex] (2,4)--(1.9999,4);\draw [-latex] (2,0)--(2.0001,0);\draw [-latex] (1.5,1.99999)--(1.5,2.0001);
\end{tikzpicture}
\caption{The contour of the $I_{m \Bar{n}}^{1\Bar{0}}$ integral.} \label{fig:I10}
    \end{center}
\end{figure}
Hence, $I_{m \bar{n}}^{1 \bar{0}} = 0$.

Alternatively, we can also compute this directly using polar coordinates in the second step of \eqref{integralresult},
\begin{equation}
    \int \frac{d^2 z}{z} = \int^{\infty}_0 d \rho \int^{2 \pi}_0 d \theta e^{- i \theta} = 0.
\end{equation}
Similarly, one has $I_{m \Bar{n}}^{0\Bar{1}} = 0$.

\section{Deformed soft graviton theorems}
\label{B}

Now we have shown that the universal IR divergence of all-loop corrected 4D amplitudes in scalar-QED and gravity are governed by the corresponding marginal deformation of 2D correlators, we would like to reformulate soft theorems \cite{Donnay:2018neh, Adamo:2019ipt, Puhm:2019zbl, Guevara:2019ypd} in the deformed CCFT.
In particular, the negative helicity soft graviton theorems can be reformulated as the following Ward identities:
\begin{equation} \label{soft}
    \< \mathcal{O}_{k, -2} X_n \>^{(0)} = \left( \frac{W^{(1-k)-}}{\epsilon} + O(\epsilon^0) \right) \<X_n\>^{(0)}, \quad k=1,0,-1,
\end{equation}
where the soft limit is recast as a limit in the conformal dimension of the soft operator $\mathcal{O}_{k, -2}$. The leading, subleading, and subsubleading soft operators have dimension $1$, dimension $0$, and dimension $-1$, respectively. The three orders of soft factors are given by
\begin{align}
    W^{(0)-} = &\sum_{k = 1}^n \frac{\Bar{z} - \Bar{z}_k}{z - z_k} \mathcal{G}_k,  \\
    W^{(1)-} = & \sum^{n}_{k=1} \frac{(z-z_k)^2}{\bar{z}-\bar{z}_k}\left( \frac{2 L_{0,k}}{z-z_k} - \p_{z_k} \right),  \\
     W^{(2)-} = & - \frac{1}{2} \sum_{k = 1}^n \frac{z - z_k}{\Bar{z} - \Bar{z}_k} \mathcal{G}_k^{-1} \bigg[ 16 L_{0,k}^2 + 4 L_{0,k} + (z-z_k)^2 \p_{z_k}^2\nn\\
    &- 4 (z - z_k) 2 \p_{z_k} L_{0,k} - 2 (z-z_k) \p_{z_k} \bigg] \label{eq:CCFTsoft}
\end{align}
where $L_{0,k}$ is the Virasoro generator, whose eigenvalue $h_k$ is the conformal weight of $k$-th operator.
Now let us consider how the flow under our deformation affects the conformally soft theorems \eqref{soft}. The corrections to the conformally soft theorem can be split into two parts:
\begin{equation}
    W^{(t)-}_{[\lambda_G]} = \widetilde{W}^{(t)-}_{[\lambda_G]} + \Delta_{\text{sub}} W^{(t)-}_{[\lambda_G]}, \quad \forall t = 0,1,2
\end{equation}
where the first term and the second term originate from the leading IR divergent term $(\frac{\sigma_n^G}{\epsilon})^{\ell} \<X_n\>^{(0)}$ and the non-universal terms $N_{\lambda_G}^{(\ell)}$, respectively. Then we will show the explicit form of $\widetilde{W}^{(t)-}_{[\lambda_G]}$ by extracting all $(\frac{\sigma_n^G}{\epsilon})^{\ell} \<X_n\>^{(0)}$ terms, denoted as $\< \widetilde{X}_n \>_{[\lambda_G]}$
\begin{equation}
    \< \widetilde{X}_n \>_{[\lambda_G]} = e^{- \frac{\lambda_G}{\epsilon} \frac{\sigma_n^G}{\epsilon}} \< X_n \>^{(0)}
\end{equation}
The multi-soft theorems ensure that the Ward identity 
remains valid even with the inclusion of additional conformally soft operators. Therefore, we can directly use the deformed correlator. 
Now we need to examine the flow of an $n+1$ point correlator where the additional particle corresponds to a soft operator,
\begin{align}
    \widetilde{W}^{(t)-}_{[\lambda_G]} \<\widetilde{X}_n\>_{[\lambda_G]} & = \epsilon \< \widetilde{\mathcal{O}_{1-t, -2} X_n} \>_{[\lambda_G]} = \epsilon e^{- \frac{\lambda_G}{\epsilon} \frac{\sigma_{n + 1}^G}{\epsilon}} \< \mathcal{O}_{1-t,-2} X_n\>^{(0)} \notag\\
    & = \epsilon e^{- \frac{\lambda_G}{\epsilon} \frac{\sigma_n^G}{\epsilon}} \sum_{k = 0}^{\infty} \frac{1}{k!} \left( - \frac{\lambda_G}{\epsilon} \frac{\sigma^{\prime G}_{n+1}}{\epsilon} \right)^k \< \mathcal{O}_{1-t+k,-2} X_n\>^{(0)} \notag \\
    & = e^{- \frac{\lambda_G}{\epsilon} \frac{\sigma^G_n}{\epsilon}} \sum_{k = 0}^t \frac{1}{k!} \left( - \frac{\lambda_G}{\epsilon} \frac{\sigma^{\prime G}_{n+1}}{\epsilon} \right)^k W^{(t-k)-} \< X_n\>^{(0)}. \label{eq:W(t)loop}
\end{align}
Note that insertions of $\mathcal{O}_{k,-2}$ vanish for $k>1$.
For notational brevity, we define
\begin{align}
    &\sigma_{n+1}^G = \sigma_n^G + \sigma^{\prime G}_{n+1} \mathcal{G}, \quad \sigma^{\prime G}_{n+1} = 2 
    \pi \sum_{i = 1}^n |z - z_i|^2 \mathcal{G}_i  \log |z - z_i|^2.
\end{align}
{ As proposed \cite{He:2017fsb} that $\Delta_{\text{sub}} W^{(t)-}_{[\lambda_G]}$ is expected to be vanishing. Consequently, to obtain the deformed soft factors $W^{(t)-}_{[\lambda_G]}$, we only need to compute the universal part $\widetilde{W}^{(t)-}_{[\lambda_G]}$ by interchanging $e^{- \frac{\lambda_G}{\epsilon^2} \sigma_n^G}$ and $W^{(t-k)-}$ in \eqref{eq:W(t)loop}. It is not necessary to distinguish $W^{(t)-}_{[\lambda_G]}$ and $\widetilde{W}^{(t)-}_{[\lambda_G]}$.} 
The commutation relations between them and the derivation of the explicit forms of the deformed soft factors are provided in the next section.
The deformed soft factors $W^{(0)-}_{[\lambda_G]}, W^{(1)-}_{[\lambda_G]}, W^{(2)-}_{[\lambda_G]}$ are exact to zeroth, first, and second order, respectively, in the expansion of the deformed parameter $\lambda_G$. They precisely recover the results of the loop-corrected soft graviton theorems presented in the following sections when the identification $\lambda_G = - \frac{\epsilon G_N}{16 \pi^2}$ is implemented.

\section{The exchange of $e^{-\frac{\lambda_G}{\epsilon^2} \sigma_n^G }$ and derivative operators}

We first compute the eigenvalues of $L_{0,k}$ and $L_{0,k}^2$ after the marginal deformations,
\begin{equation}
    \begin{split}
         L_{0,k} \<\widetilde{X}_n\>_{[\lambda_G]} & = e^{- \frac{\lambda_G}{\epsilon^2} \sigma^G_{n-1,k}} L_{0,k} \left( e^{ - \frac{\lambda_G}{\epsilon^2} \sigma^{\prime G}_{n,k} \mathcal{G}_k} \<X_n\>^{(0)} \right)\\
          &= e^{- \frac{\lambda_G}{\epsilon^2} \sigma_{n-1,k}^G} \sum_{l = 0}^{\infty} \frac{1}{l!} \left( - \frac{\lambda_G}{\epsilon^2} \sigma^{\prime G}_{n,k} \right)^l L_{0,k} \left[ (\mathcal{G}_k)^l \<X_n\>^{(0)} \right] \\
        & = e^{- \frac{\lambda_G}{\epsilon^2} \sigma_{n-1,k}^G} \sum_{l = 0}^{\infty} \frac{1}{l!} \left(h_k + \frac{l}{2} \right) \left( - \frac{\lambda_G}{\epsilon^2} \sigma'_{n,k} \right)^l (\mathcal{G}_k)^l \<X_n\>^{(0)} \\
        & = h_k \<\widetilde{X}_n\>_{[\lambda_G]} - \frac{\lambda_G}{\epsilon^2} \sigma^{\prime G}_{n,k} \mathcal{G}_k e^{ - \frac{\lambda_G}{\epsilon^2} \sigma^G_{n-1,k}} \frac{1}{2} \sum_{l = 1}^{\infty} \frac{\left( - \frac{\lambda_G}{\epsilon^2} \sigma^{\prime G}_{n,k} \right)^{l-1}}{(l-1)!} (\mathcal{G}_k)^{l-1} \<X_n\>^{(0)} \\
        & = \left( h_k - \frac{1}{2} \frac{\lambda_G}{\epsilon^2} \sigma^{\prime G}_{n,k} \mathcal{G}_k \right)\<\widetilde{X}_n\>_{[\lambda_G]},
    \end{split}
\end{equation}
and
\begin{equation}
    \begin{split}
         L_{0,k}^2 \<\widetilde{X}_n\>_{[\lambda_G]} =& e^{- \frac{\lambda_G}{\epsilon^2} \sigma_{n-1,k}^G} L_{0,k}^2 \sum_{l = 0}^{\infty} \frac{1}{l!} \left( - \frac{\lambda_G}{\epsilon^2} \sigma^{\prime G}_{n,k} \right)^l (\mathcal{G}_k)^l \<X_n\>^{(0)} \\
         = &e^{- \frac{\lambda_G}{\epsilon^2} \sigma_{n-1,k}^G} \sum_{l = 0}^{\infty} \frac{1}{l!} \left(h_k + \frac{l}{2} \right)^2 \left( - \frac{\lambda_G}{\epsilon^2} \sigma^{\prime G}_{n,k} \mathcal{G}_k \right)^l \<X_n\>^{(0)} \\
         =& e^{- \frac{\lambda_G}{\epsilon^2} \sigma_{n-1,k}^G} \sum_{l = 0}^{\infty} \frac{1}{l!} \left[h_k^2 + l \left(h_k + \frac{1}{4}\right) + \frac{l (l-1)}{4} \right] \left( - \frac{\lambda_G}{\epsilon^2} \sigma^{\prime G}_{n,k} \mathcal{G}_k \right)^l \<X_n\>^{(0)} \\
         = &h_k^2 \<\widetilde{X}_n\>_{[\lambda_G]} - \left(h_k + \frac{1}{4}\right) \frac{\lambda_G}{\epsilon^2} \sigma^{\prime G}_{n,k} \mathcal{G}_k e^{- \frac{\lambda_G}{\epsilon^2} \sigma^G_{n-1,k}} \sum_{l = 1}^{\infty} \frac{\left( - \frac{\lambda_G}{\epsilon^2} \sigma^{\prime G}_{n,k} \mathcal{G}_k \right)^{l-1}}{(l-1)!} \<X_n\>^{(0)} \\
        &\hspace{1cm}+ \frac{1}{4} \left(- \frac{\lambda_G}{\epsilon^2} \sigma_{n,k}^{\prime G} \right)^2 (\mathcal{G}_k)^2 e^{- \frac{\lambda_G}{\epsilon^2} \sigma^G_{n-1,k}} \sum_{l = 2}^{\infty} \frac{\left( - \frac{\lambda_G}{\epsilon^2} \sigma^{\prime G}_{n,k} \mathcal{G}_k \right)^{l-2}}{(l-2)!} \<X_n\>^{(0)} \\
         =& \left[ h_k^2 - \frac{\lambda_G}{\epsilon^2} \sigma^{\prime G}_{n,k} \mathcal{G}_k \left(h_k + \frac{1}{4} \right) + \frac{1}{4} \left( - \frac{\lambda_G}{\epsilon^2} \sigma^{\prime G}_{n,k} \right) (\mathcal{G}_k)^2 \right] \<\widetilde{X}_n\>_{[\lambda_G]},
    \end{split}
\end{equation}
where we have used
\begin{equation}
    \sigma_n^G = \sigma_{n-1,k}^G + \sigma^{\prime G}_{n,k} \mathcal{G}_k, \quad \sigma^{\prime G}_{n,k} = 2 r_0^2 \pi \sum_{i \neq k}^n |z_{ik}|^2 \mathcal{G}_i \log |z_{ik}|^2,
\end{equation}
and
\begin{equation}
    L_{0,k} ((\mathcal{G}_k)^l \mathcal{O}_{\Delta_k, s_k}) = \left(h_k + \frac{l}{2} \right) ((\mathcal{G}_k)^l \mathcal{O}_{\Delta_k, s_k}).
\end{equation}
Note that $L_{0,k}$ can act on $\mathcal{G}_k$, which has weights $(\frac{1}{2}, \frac{1}{2})$,
\begin{equation}
    L_{0,k} \mathcal{G}_i = \delta_{ik} \frac{1}{2} \mathcal{G}_i \Rightarrow (L_{0,k} \sigma^G_n) = \frac{1}{2} \sigma^{\prime G}_{n,k} \mathcal{G}_k.
\end{equation}
Therefore, the exchange of $L_{0,k},L_{0,k}^2$ with $e^{- \frac{\lambda_G}{\epsilon^2} \sigma_n^G}$ in 2D yields
\begin{equation} \label{eq:exchangeL}
    \begin{split}
        & h_k \<\widetilde{X}_n\>_{[\lambda_G]} = e^{- \frac{\lambda_G}{\epsilon^2} \sigma_n^G} L_{0,k} \<X_n\>^{(0)} = \Big[ L_{0,k} + \frac{\lambda_G}{\epsilon^2} (L_{0,k} \sigma_n^G) \Big] \<\widetilde{X}_n\>_{[\lambda_G]} \\
        & e^{- \frac{\lambda_G}{\epsilon^2} \sigma^G_n} (L_{0,k})^2 \<X_n\>^{(0)} = \left[ (L_{0,k})^2 + 2 \frac{\lambda_G}{\epsilon^2} (L_{0,k} \sigma_n^G) \left( L_{0,k} + \frac{1}{4} \right) + \left(\frac{\lambda_G}{\epsilon^2} L_{0,k} \sigma_n^G \right)^2 \right] \<\widetilde{X}_n\>_{[\lambda_G]},
    \end{split}
\end{equation}
where we have used
\begin{equation}
    e^{- \frac{\lambda_G}{\epsilon^2} \sigma_n^G} (L_{0,k})^l \<X_n\>^{(0)} = (h_k)^l \< \widetilde{X}_n \>_{[\lambda_G]}.
\end{equation}
Meanwhile, the exchange of $\p_{z_k},\p_{z_k}^2$ with $e^{- \frac{\lambda_G}{\epsilon^2} \sigma^G_n}$ is given by
\begin{equation} \label{eq:exchangepzk}
    e^{- \frac{\lambda_G}{\epsilon^2} \sigma^G_n} \p_{z_k} \<X_n\>^{(0)} = ( \p_{z_k} + \frac{\lambda_G}{\epsilon^2} \p_{z_k} \sigma^G_n) \<\widetilde{X}_n\>_{[\lambda_G]},
\end{equation}
and
\begin{equation}
    e^{- \frac{\lambda_G}{\epsilon^2} \sigma^G_n} \p_{z_k}^2 \<X_n\>^{(0)} = \left[ \p_{z_k}^2 + (\frac{\lambda_G}{\epsilon^2} \p_{z_k} \sigma^G_n)^2 + \frac{\lambda_G}{\epsilon^2} \p_{z_k}^2 \sigma^G_n + 2 \frac{\lambda_G}{\epsilon^2} \p_{z_k} \sigma^G_n \p_{z_k} \right] \<\widetilde{X}_n\>_{[\lambda_G]}.
\end{equation}
Finally, the exchange of $L_{0,k} \p_{z_k}$ is given by
\begin{align}
     e^{- \frac{\lambda_G}{\epsilon^2} \sigma^G_n} h_k \p_{z_k} \<X_n\>^{(0)} &= e^{- \frac{\lambda_G}{\epsilon^2} \sigma^G_n} \p_{z_k} L_{0,k} \<X_n\>^{(0)}\nn\\
    &= \left( \p_{z_k} + \frac{\lambda_G}{\epsilon^2} (\p_{z_k} \sigma_n^G) \right) \left( L_{0,k} + \frac{\lambda_G}{\epsilon^2} (L_{0,k} \sigma^G_n) \right) \<\widetilde{X}_n\>_{[\lambda_G]} \notag\\
    & = \p_{z_k} (L_{0,k} \<\widetilde{X}_n\>_{[\lambda_G]})
     + \Big[ \frac{\lambda_G}{\epsilon^2} (\p_{z_k} (L_{0,k} \sigma^G_n)) + \frac{\lambda_G}{\epsilon^2} (L_{0,k} \sigma^G_n) \p_{z_k} \notag \\
     &\hspace{2cm} + \frac{\lambda_G}{\epsilon^2} (\p_{z_k} \sigma^G_n) L_{0,k} + (\frac{\lambda_G}{\epsilon^2})^2 (\p_{z_k} \sigma^G_n) (L_{0,k} \sigma^G_n) \Big] \<\widetilde{X}_n\>_{[\lambda_G]} \notag\\
    & = \Big[ L_{0,k} \p_{z_k} + \frac{\lambda_G}{\epsilon^2} (L_{0,k} \sigma^G_n) \p_{z_k} + \frac{\lambda_G}{\epsilon^2} (\p_{z_k} \sigma^G_n) L_{0,k}\nn\\
     &+ (\frac{\lambda_G}{\epsilon^2})^2 (\p_{z_k} \sigma^G_n) (L_{0,k} \sigma^G_n) \Big] \<\widetilde{X}_n\>_{[\lambda_G]},
\end{align}
where we have used
\begin{equation}
    \p_{z_k} (L_{0,k} \<\widetilde{X}_n\>_{[\lambda_G]}) = \p_{z_k} ((h_k - \frac{\lambda_G}{\epsilon^2} (L_{0,k} \sigma^G_n)) \<\widetilde{X}_n\>_{[\lambda_G]}) = (L_{0,k} \p_{z_k} - \frac{\lambda_G}{\epsilon^2} \p_{z_k} (L_{0,k} \sigma^G_n)) \<\widetilde{X}_n\>_{[\lambda_G]}.
\end{equation}
With this, we know how to exchange $W^{(1)-}$ and $e^{- \frac{\lambda_G}{\epsilon^2} \sigma^G_n}$ past each other using \eqref{eq:exchangeL} and \eqref{eq:exchangepzk},
\begin{equation} \label{eq:exchangeW(1)}
    e^{- \frac{\lambda_G}{\epsilon^2} \sigma^G_n} W^{(1)-} \<X_n\>^{(0)} =  \left[  W^{(1)-} + \frac{\lambda_G}{\epsilon^2} (W^{(1)-} \sigma^G_n ) \right] \<\widetilde{X}_n\>_{[\lambda_G]}.
\end{equation}
Similarly, the exchange of $W^{(2)-}$ and $e^{- \frac{\lambda_G}{\epsilon^2} \sigma^G_n}$ is given by
\begin{equation}
    e^{- \frac{\lambda_G}{\epsilon^2} \sigma^G_n} W^{(2)-} \<X_n\>^{(0)} = \left[ W^{(2)-} - \frac{\lambda_G}{\epsilon^2} A_1 + (\frac{\lambda_G}{\epsilon^2})^2 A_2 \right] \<\widetilde{X}_n\>_{[\lambda_G]},
\end{equation}
where
\begin{align}
    A_1 & = - (W^{(2)-} \sigma^G_n) + \sum_{k = 1}^n \frac{z - z_k}{\Bar{z} - \Bar{z}_k} \mathcal{G}_k^{-1} \Big[ (z - z_k)^2 \p_{z_k} \sigma^G_n \p_{z_k} \nn\\
    &\hspace{2cm}- 4 (z-z_k) \Big((L_{0,k} \sigma^G_n) \p_{z_k} + (\p_{z_k} \sigma^G_n) L_{0,k} \Big) + 16 (L_{0,k} \sigma^G_n) L_{0,k} \Big] \notag \\
    & = - (W^{(2)-} \sigma^G_n) + \sum_{k = 1}^n \frac{z - z_k}{\Bar{z} - \Bar{z}_k} \mathcal{G}_k^{-1} \Big[ (z - z_k) (\p_{z_k} \sigma^G_n) - 4 (L_{0,k} \sigma^G_n) \Big]
    \Big[ (z - z_k) \p_{z_k} - 4 L_{0,k} \Big], \label{eq:exchangeW(2)A1}
\end{align}
and
\begin{equation} \label{eq:exchangeW(2)A2}
    \begin{split}
        A_2 & = - \frac{1}{2} \sum_{k = 1}^n \frac{z - z_k}{\Bar{z} - \Bar{z}_k} \mathcal{G}_k^{-1} \Big[ (z - z_k)^2 (\p_{z_k} \sigma^G_n)^2
         - 8 (z - z_k) (\p_{z_k} \sigma^G_n) (L_{0,k} \sigma^G_n) + 16 (L_{0,k} \sigma^G_n)^2 \Big] \\
        & = - \frac{1}{2} \sum_{k = 1}^n \frac{z - z_k}{\Bar{z} - \Bar{z}_k} \mathcal{G}_k^{-1} \Big[ (z - z_k) (\p_{z_k} \sigma^G_n) - 4 (L_{0,k} \sigma^G_n) \Big]^2.
    \end{split}
\end{equation}

\section{Deformed soft factors}
\label{C}

The subleading soft graviton theorem in celestial holography corresponds to the CCFT's stress tensor \cite{Kapec:2016jld}.
Loop corrections to the subleading soft graviton theorem induce a transformation in the CCFT's stress tensor, providing further validation of our consistency \cite{He:2017fsb}. Additionally, as a stringent test of our framework, we have verified that the deformed Ward identity of the stress tensor, emerging as a shadow transformation of the deformed subleading soft theorem, precisely aligns with the amplitude-based derivation \cite{He:2017fsb}.

Here we present the details for the computation of the deformed soft factor. The case $t = 0$ yields the leading deformed soft factor,
\begin{equation}
    W^{(0)-}_{[\lambda_G]} \< \widetilde{X}_n \>_{[\lambda_G]} = \epsilon \<\widetilde{\mathcal{O}_{1,-2} X_n}\>_{[\lambda_G]} = W^{(0)-} \< \widetilde{X}_n \>_{[\lambda_G]}.
\end{equation}
So we see that the leading order result does not flow. This is the dual interpretation of the fact that the leading soft graviton theorem does not receive loop corrections \cite{Weinberg:1965nx}.

The case $t = 1$ yields the flow effect of subleading soft theorem,
\begin{equation} \label{correctedsub}
    \begin{split}
        W^{(1)-}_{[\lambda_G]} \<\widetilde{X}_n\>_{[\lambda_G]} & = \< \mathcal{O}_{0, -2} \widetilde{X}_n \>_{[\lambda_G]} = e^{- \frac{\lambda_G}{\epsilon^2} \sigma^G_n} \left[W^{(1)-} - \frac{\lambda_G}{\epsilon^2} \sigma^{\prime G}_{n+1}W^{(0)-} \right] \< X_n\>^{(0)} \\
        & = \left[ W^{(1)-} - \frac{\lambda_G}{\epsilon^2} \left(\sigma^{\prime G}_{n+1} W^{(0)-} - (W^{(1)-} \sigma^G_n) \right) \right] \< \widetilde{X}_n\>_{[\lambda_G]}.
    \end{split}
\end{equation}
By using the identification $\lambda_G = - \frac{\epsilon G_N}{16 \pi^2}$, which has been introduced in the main text, the above equation recovers precisely the all loop corrected subleading soft graviton theorem \cite{Bern:2014oka,He:2014bga}.

The case $t = 2$ yields the flow effect on the subsubleading soft theorem,
\begin{equation}
    \begin{split}
        W^{(2)-}_{[\lambda_G]} \<\widetilde{X}_n\>_{[\lambda_G]} & = \< \widetilde{ \mathcal{O}_{-1, -2} X_n } \>_{[\lambda_G]}\\
         &= e^{-\frac{\lambda_G}{\epsilon^2} \sigma_n^G} \left[ W^{(2)-} - \frac{\lambda_G}{\epsilon^2} \sigma^{\prime G}_{n+1} W^{(1)-} + \frac{1}{2} (\frac{\lambda_G}{\epsilon^2} \sigma_{n+1}^{\prime G})^2 W^{(0)-} \right] \< X_n\>^{(0)} \\
        & = \left[ W^{(2)-} + \lambda W^{(2)-}_{\lambda^1} + \lambda^2 W^{(2)-}_{\lambda^2} \right] \<\widetilde{X}_n\>_{[\lambda_G]},
    \end{split}
\end{equation}
where
\begin{equation} \label{eq:2Dsoft(2)}
    \begin{split}
        W^{(2)-}_{\lambda^1} = & \sigma^{\prime G}_{n+1} W^{(1)-} - (W^{(2)-} \sigma^G_n)  \\&\hspace{0.5cm}
        + \sum_{k = 1}^n \frac{z - z_k}{\Bar{z} - \Bar{z}_k} \mathcal{G}_k^{-1} \Big[ (z - z_k) (\p_{z_k} \sigma^G_n) - 4 (L_{0,k} \sigma^G_n) \Big] \Big[ (z - z_k) \p_{z_k} - 4 L_{0,k} \Big], \\
        W^{(2)-}_{\lambda^2} = & \frac{(\sigma_{n+1}^{\prime G})^2}{2} W^{(0)-} - \sigma^{\prime G}_{n+1} (W^{(1)-} \sigma^G_n)  \\&\hspace{2cm}
        - \frac{1}{2} \sum_{k = 1}^n \frac{z - z_k}{\Bar{z} - \Bar{z}_k} \mathcal{G}_k^{-1} \Big[ (z - z_k) (\p_{z_k} \sigma^G_n) - 4 (L_{0,k} \sigma^G_n) \Big]^2 .
    \end{split}
\end{equation}
Since $\Delta_{\text{sub}} W^{(t)-}_{[\lambda_G]}$ is expected to be vanishing, this recovers the amplitude result for the subsubleading soft theorem in \cite{Bern:2014oka}.

\section{Shifted stress tensor}
\label{E}

As a further consistency check of our proposal, it can be demonstrated that the subleading soft graviton theorem in the deformed correlator recovers the loop-corrected stress tensor of the corresponding CCFT \cite{Kapec:2016jld}. The 2D stress tensor with dimension $\Delta=2$ is the shadow transformation of the subleading soft-graviton operator with dimension $\Delta=0$ \cite{Kapec:2017gsg}:
\begin{equation}
T_{ww}=\frac{3!}{4\pi} \int \td^2 z \frac{1}{(w-z)^4} \epsilon \mathcal{O}_{0,-2}(z).
\end{equation}
The subleading soft graviton theorem receives loop corrections \cite{Bern:2014oka, He:2014bga}, which correspond to a shift in the CCFT stress tensor \cite{He:2017fsb} (see also \cite{Donnay:2022hkf, Pasterski:2022djr}). Here, we will demonstrate how the stress tensor is shifted purely from a CCFT point of view. By implementing a shadow transformation on the deformed subleading soft theorem presented explicitly in \eqref{correctedsub}, the deformed Ward identity of the stress tensor can be derived as follows:
\begin{align}
    \<T_{ww} X_n\>_{[\lambda_G]} \nn
    =&\frac{3!}{4\pi} \int \td^2 z \frac{1}{(w-z)^4} \epsilon \<\mathcal{O}_{0,-2} X_n\>_{[\lambda_G]} \nn\\
    =&-\frac{1}{4\pi}\int \td^2 z \frac{1}{(w-z)}
    \partial_z^3\bigg[ W^{(1)-} - \frac{\lambda_G}{\epsilon^2} \left( \sigma^{\prime G}_{n+1} W^{(0)-}- (W^{(1)-} \sigma^G_n) \right) \bigg] \< X_n\>_{[\lambda_G]}.
\end{align}
We see that the stress tensor is shifted by $\tilde{T}_{ww}={T}_{ww}-\Delta {T}_{ww}$, where
\be
    \<\Delta T_{ww} X_n\>_{[\lambda_G]}
    = \frac{1}{4\pi} \frac{\lambda_G}{\epsilon^2} \int \td^2 z \frac{1}{w-z}
     \partial_z^3\left( \sigma^{\prime G}_{n+1} W^{(0)-}- (W^{(1)-} \sigma^G_n) \right)  \< X_n\>_{[\lambda_G]}.
\ee
The shifted stress tensor $\tilde{T}_{ww}$ obeys the standard Ward identity of a stress tensor for a CFT. This result is in exact agreement with the derivation from the amplitude side \cite{He:2017fsb}.


\providecommand{\href}[2]{#2}\begingroup\raggedright\endgroup

\end{document}